\documentclass[preprint2]{aastex631}
\usepackage{graphicx,amsmath,amsthm}

\newcommand \teff {T_{\rm eff}}

\shorttitle{Brightness Fluctuation Spectra of Sun-Like Stars}
\shortauthors{Brown, Garc\'\i a, Mathur, Metcalfe, \& Santos}

\begin{document}

\title{Brightness Fluctuation Spectra of Sun-Like Stars. I. The Mid-Frequency Continuum}

\author[0000-0001-5062-0847]{Timothy M.~Brown}
\affil{Las Cumbres Observatory, Goleta CA 93117, USA}
\affil{CASA, University of Colorado, Boulder, CO 80309, USA}
\author[0000-0002-8854-3776]{Rafael A.~Garc\'\i a}
\affil{AIM, CEA, CNRS, Universit\'e Paris-Saclay, Universit\'e Paris Diderot, Sorbonne Paris Cit\'e, F-91191 Gif-sur-Yvette, France}
\author[0000-0002-0129-0316]{Savita Mathur}
\affil{Instituto de Astrof\'{i}sica de Canarias, 38200 La Laguna, Tenerife, Spain}
\affil{Universidad de La Laguna, Dpto. de Astrof\'{i}sica, 38205 La Laguna, Tenerife, Spain}
\author[0000-0003-4034-0416]{Travis S.~Metcalfe}
\affil{White Dwarf Research Corporation, 9020 Brumm Trail, Golden CO 80403, USA}
\author[0000-0001-7195-6542]{\^Angela R.~G.~Santos}
\affil{Department of Physics, University of Warwick, Coventry, CV4 7AL, UK}

\begin{abstract} 
We analyze space-based time series photometry of Sun-like stars, mostly in the Pleiades,
but also field stars and the Sun itself.
We focus on timescales between roughly 1 hour and 1 day.
In the corresponding frequency band 
these stars display brightness fluctuations with a decreasing 
power-law continuous spectrum.
\textit{K2} and \textit{Kepler} observations
show that
the RMS flicker due to this Mid-Frequency Continuum (MFC)
can reach almost 1\%, approaching 
the modulation amplitude from active regions. 
The
MFC amplitude varies by a factor
up to 40 among Pleiades members with similar $\teff$,
depending mainly on the stellar 
Rossby number $Ro$.  
For $Ro \leq 0.04$, the mean amplitude 
is roughly constant at about 0.4\%;
at larger $Ro$ the amplitude decreases rapidly, shrinking by about two orders of magnitude for $Ro \simeq 1$.  
Among stars, the MFC amplitude correlates poorly
with that of modulation from rotating active regions.
Among field stars observed for 3 years by \textit{Kepler}, the 
quarterly average modulation amplitudes from active regions are much more
time-variable than the quarterly MFC amplitudes.
We argue
that the process causing the MFC is largely magnetic
in nature,  
and that its power-law spectrum comes from
magnetic processes distinct from
the star's global dynamo, with shorter timescales.
By analogy with solar phenomena, we hypothesize that the MFC
arises from a (sometimes energetic) variant of the solar 
magnetic network,
perhaps combined with rotation-related changes in the morphology of supergranules.
\end{abstract}

\keywords{Solar dynamo; Stellar activity; Stellar rotation; Supergranulation}

\NewPageAfterKeywords

\section{Introduction}\label{sec1}
It has long been known that some stars have surface magnetic
active regions (spots and faculae) analogous to those seen on the Sun,
\citep[e.g.][]{kron1947, kron1952, chugainov1966, bopp1973}.
The principal evidence for such starspots is the periodic variation of
stellar brightness as stellar rotation carries the active
regions across the star's visible hemisphere. 

With the advent of spaceborne time-series photometry, it became possible to
detect active regions even on stars with roughly solar activity levels.
In the last 15 years, missions including \textit{MOST} (Microvariability \& Oscillations of STars; Matthews et al. 2004), \textit{CoRoT} (Convection, Rotation, and Transits; \citet{baglin2006}),  \emph{Kepler} \citep{borucki2010}, \textit{K2} \citep{howell2014}, and TESS \citep[Transiting Exoplanet Survey Satellite;][]{ricker2014}
have produced stellar photometry with precision and temporal coverage that is not possible from the ground.
In particular, the \textit{Kepler} and \textit{K2} missions 
generated prolonged and precise time series of hundreds of thousands of stars,
allowing us to characterize in detail the brightness signatures of stellar
activity modulated by rotation
\citep{garcia2014, mcquillan2014, santos2019, gordon2021}.
Among the available mission results, those from \textit{K2} stand out because some
\textit{K2} campaigns pointed at nearby open star clusters.
Exploiting the data from these campaigns gives all of the usual dividends
expected from open cluster data,
allowing unambiguous comparisons among member stars.

Here we discuss mostly observations of the Pleiades obtained by the \textit{K2} mission, to investigate brightness fluctuations with timescales
between roughly a day and the \textit{K2} long-cadence Nyquist period of about an hour.
These timescales are longer than granulation lifetimes,
but shorter than the great majority of stellar rotation periods,
or the lifetimes of magnetic active regions.
The corresponding frequencies lie between roughly 20 and 300 $\mu$Hz.
To align with the \textit{K2} long-cadence frequency coverage, we will usually confine our interest to a slightly
smaller frequency range, $viz$ 20 $\leq \nu \leq$ 285 $\mu$Hz.
\citet{rebull2016, stauffer2016} and \citet{rebull2016a} studied the \textit{K2} time series of the Pleiades
in great detail, producing (among many other things) a list of 759 cluster
members with reliably measured rotation periods, covering all spectral types
from B to M.
Here we study mostly stars drawn from this list.

\begin{figure*}[t]      
 \centering\includegraphics[angle=90,width=5.5in]{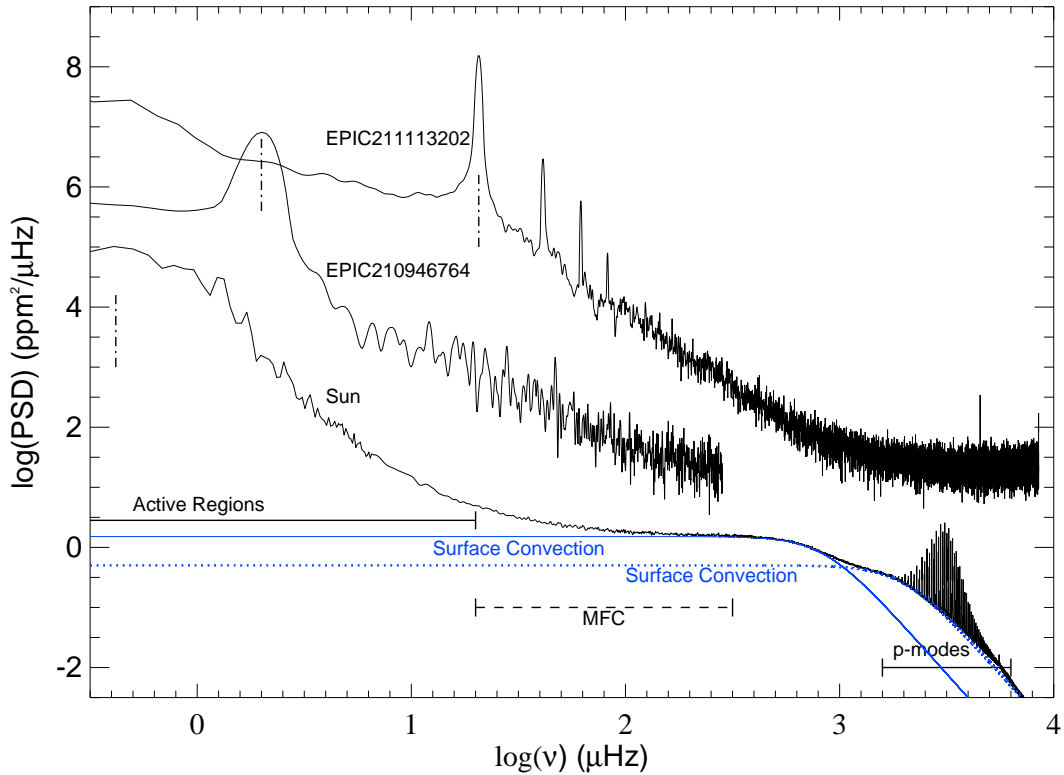} 
 \caption{Power spectra of fluctuations in the relative intensity in the
visible-light {\textit{Kepler}} bandpass, measured for two members of the Pleiades
star cluster and the Sun. Note logarithmic scales on both axes, and also that all 
3 stars are displayed on the same vertical scale; they are $not$ arbitrarily displaced
along the $y$-axis for clarity.
Annotations show the frequency ranges over which various physical processes
on the Sun are thought to be important in the spectrum, as described in the text.
The blue and blue-dotted curves indicate components of the spectrum that
are commonly fit in the literature using Harvey functions;
together, they account for the solar granulation.
The black dashed horizontal bar shows the frequency range
dominated by the Mid-Frequency Continuum (MFC) discussed in this work.
Above about $10^3$ $\mu$Hz, the flat spectrum of EPIC211113202 is from
photon shot noise.\label{fig1}}
 \end{figure*} 

Figure~\ref{fig1} shows power spectra of relative brightness fluctuations for two 
Pleiades members (EPIC211113202, $\teff$ = 4843K, mass=0.764 $M_\sun$; EPIC210946764, $\teff$=5547K, mass=0.974 $M_\sun$)
and the Sun ($\teff$ = 5777K, mass=1 $M_\sun$).
The upper spectrum comes from \textit{K2} short-cadence data (SC = 1 minute sampling),
and the middle one from \textit{K2} long-cadence data
(LC = 1/2 hour sampling).
The solar spectrum is from a 23-year time series of data from VIRGO/SPM \citep{frohlich1995, frohlich1995a, jimenez2002}
taken  with 1-minute observing cadence, starting near solar minimum in 1996. 

A striking feature of Figure~\ref{fig1} is the wide range of variability amplitude
observed on otherwise fairly similar stars.
The three shown here are all dwarfs
with vigorous surface convection zones,
and masses that differ by at most 25\%.
Yet among the three stars plotted here, the photometric
variability amplitude ranges over a factor of about 200.
The property that chiefly distinguishes these three stars from one another
is their rotation period $P_\text{rot}$, which varies among these stars by
a factor of about 100.
As we shall see below, rotation is key to understanding the variability
of our stellar sample.

It is useful to decompose the stellar relative brightness power spectra into two components.
First, there are periodic variations that remain coherent over time spans that
cover at least a sizable fraction of the total duration of the data set.
Physically, these arise from long-lived localized brightness features 
(presumably spots and faculae) that
persist for a few to many rotation periods.
These cause narrow peaks in the temporal power spectrum, at or near integer
multiples of the star's rotation frequency.

Second, there are variations that are incoherent over time spans of a 
few rotation periods or less.
These lead to continuous power spectra.
For a large majority of the stars
we have examined, the signal from the incoherent variations appears in
the spectra as
power laws, with the power spectral density ($PSD$) given by
$PSD = A \nu^{-\alpha}$, valid in different stars over various
ranges of $\nu$.
We call particular attention to the dominant
role of such continua in all the stars shown in
Figure~\ref{fig1},
for frequencies between the long-cadence observations' Nyquist frequency (about 285 $\mu$Hz) and
some minimum frequency $\nu_\text{min}$, where for the
purposes of this paper we will always take
$\nu_\text{min} =$ 20 $\mu$Hz.
This frequency range is indicated by the bar labeled ``MFC''
at the
bottom of the plot.

Note that for main-sequence stars,
oscillations, 
granulation, and related flows 
described in the p-mode literature (\citet{kallinger2014, santos2018}, and references therein) have 
timescales that are short
compared to the \textit{K2} long-cadence sampling time.
Hence, within the frequency range marked ``MFC" in Figure~\ref{fig1}, the high-frequency processes  contribute only
a near-constant power background.

Although the mid-frequency power-law continua are prominent in the brightness fluctuation spectra of most dwarf stars, they
have not yet been studied in a systematic way, except as a noise source
with which planetary transit observations must contend \citep{sulis2020}.
The current paper is intended as the first in a series that will
study their properties and physics.

The rest of this paper is laid out as follows:
Section~\ref{sec2} turns briefly to the spectrum of the Sun's photometric fluctuations,
to provide context for the following discussion of distant stars.
Section~\ref{sec3} describes the \textit{K2} observations, how we chose our sample of Pleiades
stars,
the model that we use to describe
their power spectra, and our techniques for parameter fitting.  Section~\ref{sec4}
examines the relationships that emerge among the stellar structure 
properties and the various fitted spectrum model parameters for our sample.
Section~\ref{sec5} considers whether the MFC is
merely an artifact of the stellar global dynamo,
by appealing to data from
sources other than \textit{K2}, and from stars that are not members of the Pleiades
cluster.
Section~\ref{sec6} attempts a coherent description of all these results, ending with a consistent but speculative interpretation
of the power-law continua in terms
of a variant of the solar magnetic network --  one that brightens
dramatically with faster stellar rotation,
and that may involve rotation-dependent changes in supergranule morphology.

\section{The Sun's Photometric Fluctuations}\label{sec2}

Our knowledge of the photometric variability spectrum of the Sun (Figure~\ref{fig1}) can help
us to understand the variability spectra of other stars.
As others have noted \citep{ulrich1970, harvey1985, garcia2009, karoff2012}, in this spectrum
one can identify frequency ranges in which
one or two physical processes are the dominant sources of variability;
several of these are shown in Figure~\ref{fig1}.
At high frequencies (above 1000 $\mu$Hz, say), the dominant sources of
variability are almost purely dynamical (granulation and p modes).
At the lowest frequencies (below roughly 20 $\mu$Hz), the variations are either essentially
magnetic (growth and decay of magnetic active regions powered by a
global dynamo), or
else arise from rotation of active regions across the solar disk.

At intermediate frequencies, there is evidence that the fluctuating sources
have both dynamical and magnetic properties.
For instance, facular elements and photospheric bright points seem to be created when granular and supergranular flows,
permeated and constrained by magnetic fields, generate evacuated flux tubes
having strong radiative signatures.
But at the applicable spatial and temporal scales for these short-lived flows, 
it is doubtful
that the processes at work connect in any direct way to the Sun's global dynamo.

These considerations raise questions about the MFC frequency range identified with the dashed bar in Figure~\ref{fig1},
spanning roughly
20 $\leq \nu \leq$ 300 $\mu$Hz.
Is the power in this region a symptom of the global dynamo?
Is the underlying mechanism essentially fluid dynamic, or do magnetic
fields play an important role?
Is the power in the MFC related to known near-surface
phenomena on the Sun, such as facular regions, granulation, or supergranulation?

\begin{figure}[t]
 \centering\includegraphics[angle=90,width=\columnwidth]{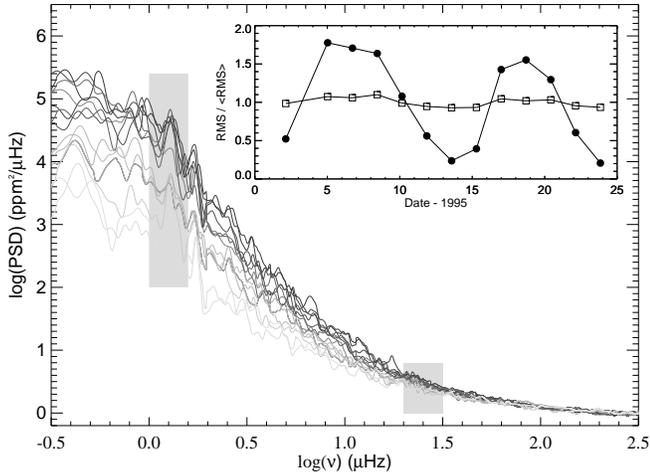} 
 \caption{Solar relative brightness fluctuation power spectra $vs.$ $\nu$, for each of the
time intervals (blocks) described in the text, shaded according to the RMS fluctuation amplitude for the block;
darker shading indicates greater RMS.
The sub-panel at upper right shows the relative RMS for each block, constrained to the 
limited frequency ranges
\{1.0 to 1.6\} $\mu$Hz (solid circles) and \{20 to 32\} $\mu$Hz (open squares).
These frequency ranges are shown as shaded bars on the main plot.
\label{fig2}}
 \end{figure}

To address some of these questions, it is instructive to examine how the 
power spectrum of the Sun's fluctuations varies with the solar cycle.
The VIRGO/SPM time series whose spectrum appears in Figure~\ref{fig1} spans two full solar
cycles.
We have broken the complete time series into 13 almost-contiguous blocks, each 625 days
in length, and computed the power spectra of these blocks individually.
In Figure~\ref{fig2} we overplot the smoothed power spectra of these blocks,
illustrating that the block-to-block solar photometric variability is much larger at
very low frequencies (below 20 $\mu$Hz) --  identified with ``Active Regions''
in Figure~\ref{fig1}) than at higher frequencies.
The typical $PSD$ values at the lowest frequencies vary by
up to 1.5 orders of magnitude during the cycle. 
Moreover, the variability is modulated in phase with the solar sunspot cycle.
We infer that the low-frequency intensity fluctuations  arise from 
processes that participate in
the global dynamo cycle, e.g., evolution of magnetic active regions,
and modulation of their photometric signature by rotation.

For frequencies above about 20 $\mu$Hz, however, the fluctuating brightness
power shows no such obvious cycle dependence;
at these frequencies the cyclic variation of the observed power is much
smaller, with block-to-block root mean square (RMS) variability of only a few percent.
This suggests that the mechanism driving photometric fluctuations
in the intermediate frequency range
is not closely connected to the global dynamo.
The relevant mechanisms may or may not involve magnetic fields in an essential
way (we will argue later that they do), but in any case they do not arise
directly from the conventional manifestations of the Sun's magnetic
cycle.

At higher frequencies ($\nu \ge$ 300 $\mu$Hz, approximately) solar photometric
fluctuations have generally been modeled using Harvey functions \citep{harvey1985}
\begin{equation}
PSD_{\rm Harvey} \ \propto \ 1/(1\ + \ (\nu / {\nu}_0 )^n ) \ ,
\end{equation}
where $\nu_0 = (2 \pi \tau)^{-1}$, and $\tau$ is a characteristic timescale
(e.g., \citet{mathur2011, kallinger2014} and \citet{santos2018}, their Figure~2).
Thus, for frequencies below $\nu_0$, the Harvey functions are nearly constant, 
independent of $\nu$.
Two such Harvey functions combine in the solar spectrum to describe the solar granulation.
These have fitted timescales $\tau \ \simeq \{ 80, 190 \}$~s,
and make roughly equal contributions of about
35 ppm RMS to the solar variability.
There is little consensus in the literature
as to what these components should be called,
so in Figure~\ref{fig1} we term them both ``surface convection".

If we take the Harvey-function models seriously, we conclude that
the solar power spectrum shows unexplained excess power, largest at the
low frequencies, in $20 \leq \nu \leq 300\ \mu$Hz.
Since we do not yet understand the physical nature of this excess, for naming
purposes we revert to observational morphology, and for now
we term it the Mid-Frequency Continuum (MFC).
If needed, a more physics-related name can be applied in due course.

The solar MFC contributes about 25 ppm RMS to the Sun's photometric
variability in a broad wavelength bandpass (Green plus Red,
similar to the 400 to 850 nm bandpass used by the $Kepler$ mission) -- an amount
that is similar in magnitude to the disk-averaged signals from p-modes or granulation, but much smaller than that caused
by solar magnetic active regions.
In what follows, we investigate the behavior of this feature
as seen in distant stars, notably in the Pleiades cluster.
We will show that in younger, more rapidly rotating stars, the 
amplitude of the MFC can be much greater than it is in the Sun.

\section{Target Sample, Data Preparation, Derived Indices}\label{sec3}

NASA's \textit{K2} mission targeted M45 (the Pleiades open cluster) during \textit{K2}
Campaign 4, running from 2015 Feb. 07 to 2015 Apr. 23 \citep{howell2014} \footnote{doi:10.17909/T9N889}.

\subsection{Star Sample and Stellar Properties}\label{Sect3.1}

We selected 101 stars to be analyzed from the list of 759 Pleiades stars with
reliable rotation periods from \citet{rebull2016} (henceforth R16).
Our aims were to select a subsample of the R16 stars (to keep the manual
labor associated with our data analysis methods within bounds), while 
achieving  adequate statistics for stable conclusions, and also providing
fairly constant numbers of stars in color bins 
0.5 magnitude wide, spanning 0.5 $\le V-K \leq$ 5.0.
This selection is reasonably complete for stars with $V-K \leq$ 2.5,
becoming increasingly incomplete for redder colors.
For example, in R16 the magnitude bin 4.0 $\leq V-K \leq$ 4.5 contains
34 stars, whereas our sample contains only 4.

For our sample of 101 stars, we took rotation periods $P_{\rm rot}$
and membership assessments (``not member'', ``good'', ``best'') from R16.
We obtained stellar classification estimates ($\teff$, $\log(g)$,
$v\sin i$) from the APOGEE DR16 ASPCAP pipeline \citep{ahumada2020, jonsson2020}.
Three of our selected stars did not have APOGEE data;  for those we used
parameter values from the MAST EPIC catalog \citep{huber2016} \footnote{https://archive.stsci.edu/k2/epic/search.php}.
Stellar magnitudes $V$ and $K$, and $V-K$ colors, came from R16.
We estimated stellar radii by combining measured values of $K$ magnitude, $\teff$, and Gaia parallax. For 24 stars, not all of these properties were available;  in these cases we used radius estimates from the MAST EPIC catalog \citep{huber2016}.
To identify binary or multiple stars, we compared the position of each star
on a ($V$ $vs$ $V-K$) color-magnitude diagram to the position of the
single-star locus, which we fitted using photometry from R16.
We identified stars more than \{0.3, 0.45\} magnitudes brighter than the locus as
possible or probable binaries, respectively.
For the Pleiades members in our star sample, the RMS
scatter in parallax is about 2\% of the mean parallax.
Thus, we do not expect the finite cluster depth to have
a significant effect on our photometric binary classifications.
The stellar power spectra also sometimes contained evidence for binarity.
In these cases the spectra displayed two or more harmonic sequences of
coherent periodic fluctuations.
If the ratio of the fundamental periods of these sequences was less than
1.4, we considered them to be examples of single-star differential rotation
\citep{donahue1996};
if greater than 1.4, we deemed them probable binaries.

In the final list, 6 stars are identified as non-members of the Pleiades.
We retained those stars to allow a preliminary assessment of differences,
if any, between members and (presumably older) field stars.
24 of our stars are classified as probable binaries, and 8 are possibly
binary.

We obtained EVEREST \citep[EPIC Variability Extraction and Removal for Exoplanet Science Targets,][]{luger2016} Long-Cadence (LC) time series for each of 
our sample stars from the Mikulski Archive for Space Telescopes (MAST\footnote{https://archive.stsci.edu/hlsp/everest}). For each time series we post-processed the data following the procedures explained in \citet{garcia2011} and we interpolated the gaps following inpainting techniques using a multiscale discrete cosine transform \citep{garcia2014a, pires2015}.
The gap-filled time series spanned 71.8 days with a mean sampling cadence of
29.42 minutes, for a total of typically 3465 samples per star.

\subsection{Short-Cadence Observations from \textit{Kepler}}

To extend our star sample to older, more slowly rotating stars, 
to see how their photometric power spectra varied on few-year timescales,
and to verify that \textit{Kepler} Short-Cadence (SC) observations yielded results
similar to those from LC observations, we analyzed
a selected group of 11 stars from the \textit{Kepler} field.
For this purpose, we chose SC stars that had been analyzed with 
asteroseismology \citep{lund2017}, and that also have been the subject
of careful studies of their rotation and magnetic activity \citep{garcia2014, santos2018, santos2019}. 
Hence these stars have accurate estimates of mass, $\teff$, radius, and $P_\text{rot}$.
In hopes of learning something about stellar activity cycles,
we preferred stars with evidence of time-varying magnetic activity \citep{mathur2014, santos2018}.
As a consequence of our selection strategy, this group of stars is hotter on
average than the Pleiades sample, and also older, with slightly lower
surface gravity.

Structural parameters for these stars come from \citet{silvaaguirre2017}, except for KIC3733735, which is from \citet{mathur2014}.
We processed \textit{Kepler} SC data following the methods described in \citet{garcia2011} (the so-called KEPSEISMIC series\footnote{https://archive.stsci.edu/prepds/kepseismic/}). As for the K2 time series, the gaps were filled with the inpainting algorithm mentioned earlier.
The time series coverage shows small differences from star to star, but
typically each contains about 1.7$\times 10^6$ samples spanning 3.14 years
(usually \textit{Kepler} quarters Q5 to Q16),
with a mean time between samples of 58.85s.

\subsection{Power Spectrum Fitting -- Definition of the MFC}

From the time series of relative brightness for each star,
we computed the power spectral density ($PSD(\nu)$, where 
$PSD$ has units of ppm$^2/\mu$Hz).
We then fit each power spectrum with a model
\begin {equation}
PSD(\nu)\  = \ P_\text{cont}(\nu) \ + P_\text{harm}(\nu) .
\end{equation}
Here $P_\text{cont}$ is a two-part broken power law function of $\nu$,
and $P_\text{harm}$ is a sum of generalized Lorentzian line profiles with frequencies that are integer multiples of the star's rotation frequency.
The power-law index for $P_\text{cont}$ is allowed to be discontinuous across the break frequency $\nu_0$
but constant on either side of it, and
the two power-law amplitudes are constrained such that $P_\text{cont}$ is continuous across the break.
For all of the fits discussed in this work,
the break frequency $\nu_0$ was arbitrarily fixed at $20 \mu$Hz.
The parameters describing the model $PSD(\nu)$ are then the amplitude $A$ of the continuum model at the break frequency $\nu_0$, the
power-law indices $\alpha_1, \alpha_2$ that apply respectively above and below $\nu_0$,
and the list of Lorentzian frequencies, amplitudes, and widths that define $P_\text{harm}(\nu)$.

For a more detailed description of how we carried out the power spectrum computation and  model fitting, see Appendix~\ref{appa}.

\subsection{Fitted Parameters}

Our fitting process yields three derived quantities that are important in 
the following discussion, as well as a number of others that we treat
as nuisance parameters.
The important quantities are as follows.

The total harmonic RMS, denoted $\sigma_\text{H}$, is the RMS photometric variability
(ppm) attributable to all frequencies less than $\nu_0$, plus any
portion of $P_\text{harm}$, the fitted harmonic power, that lies at $\nu \geq \nu_0$.
It represents the variability due to rotating active regions, and from any
other sources with periods longer than about half a day.
This parameter is closely similar to the photometric
magnetic activity index  $S_{\rm ph}$ defined by
\citet{mathur2014}.
$\sigma_{\text(H)}$ is essentially a measure of
the spottedness of the stellar photosphere, and hence
is one index of the general level of magnetic activity.
Other activity indices exist ($e.g.$, R-values computed from
equivalent widths of H$\alpha$ or the CaII infrared
triplet), which typically show behavior that is similar to that of 
spottedness, but differs in detail.
It would be interesting to trace the
connections between these various indices and the broadband
photometric parameters we describe below,
but this is beyond the scope of the present
work.

The continuum RMS, denoted $\sigma_\text{C}$, is the RMS photometric variability (ppm)
attributable to the Mid-Frequency Continuum.
We calculate it from the model $PSD$, excluding from the integral 
frequencies below $\nu_0$ and above $\nu_\text{max}$:
\begin{equation}
\sigma_\text{C} \ = \ \left[ \int_{\nu_0}^{\nu_\text{max}} P_\text{cont}(\nu) d\nu \right]^{1/2} \ .
\end{equation}

The MFC slope $\alpha_1$ is $d \log (P_\text{cont} ) / d \log \nu$ in the power-law 
domain of the photometric fluctuating power spectrum.

The lowest harmonic frequency $\nu_\text{H1}$ is less
important than the other parameters, but it is occasionally
useful.
It is the fitted frequency of the
fundamental harmonic peak in the $PSD_\text{harm}$ spectrum.
Absent a high-quality estimate from the literature, we sometimes use this
value to compute a star's rotation period as 
$P_\text{rot} = 11.574\ \mu {\rm Hz} / \nu_\text{H1}$ days.

We have applied our fitting procedure to the list of 101
selected \textit{K2} Campaign 4 (``Pleiades'') stars, the 11 \textit{Kepler}
short-cadence stars, and the Sun.
Table~\ref{tab1} contains the physical characteristics of all of these stars from sources in the literature as described above, as well as our measured values for the three main fitted parameters 
just described.
Other parameters that emerge from the fits (e.g., the slope of the 
low-frequency
continuum, the periods and linewidth parameters of the higher-order
harmonic peaks) may perhaps contain useful information, but they do not
enter into our current discussion.  

\section{Dependence of MFC on Stellar Parameters}\label{sec4}

The MFC amplitude $\sigma_\text{C}$ is highly variable across stars;
observed values in our sample range over
almost 3 orders of magnitude.
The same is true of the global activity index
$\sigma_\text{H}$.
And yet, importantly, these two quantities are
only modestly well correlated:
across our full star sample, the Spearman rank correlation between $\log(\sigma_\text{H})$ and $\log(\sigma_\text{C})$ is only 0.57.
Though this correlation is highly significant,
it is not good enough to allow using one of these amplitudes
to make a meaningful prediction of the other.
Thus, near any given value of $\log(\sigma_\text{H})$,
the range of $\log(\sigma_\text{C})$ is
typically 2 dex.
What, then, do these amplitudes depend on?
We first address $\sigma_\text{C}$,
because in this case the answer is fairly clear.
In a later section we turn to the question
of whether $\sigma_\text{C}$ and $\sigma_\text{H}$ may be considered
as different manifestations of a single process.

 \begin{figure*}[t]
 \centering\includegraphics[angle=90,width=5.5in]{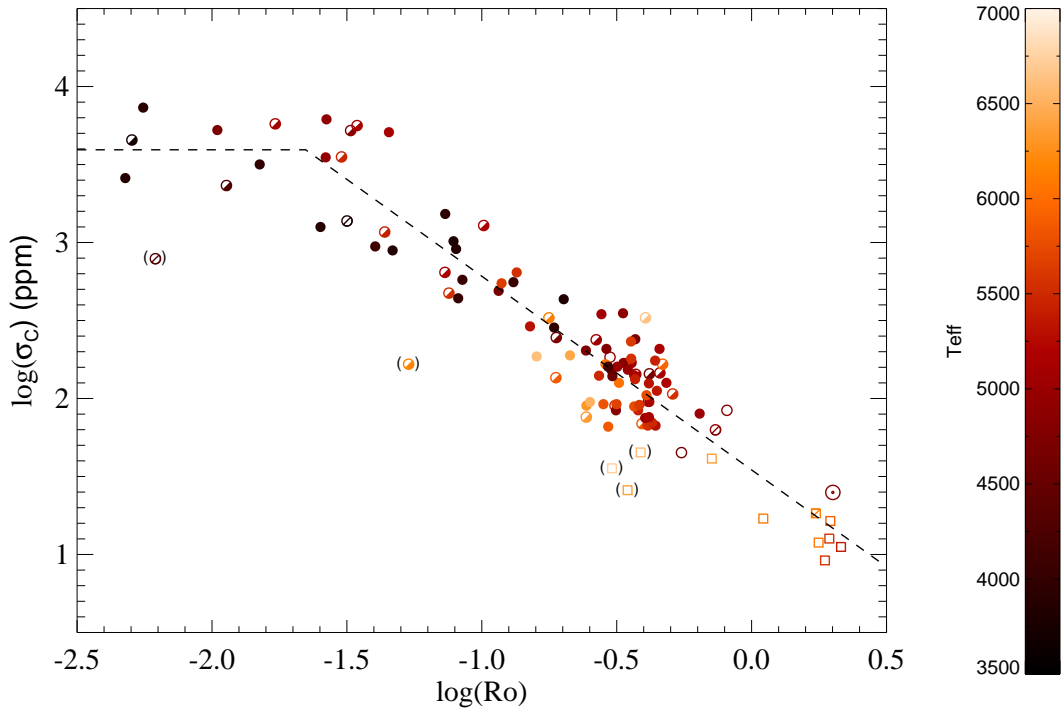} 
 \caption{$\log \sigma_\text{C}$ $vs$ $\log(Ro)$, for all stars in our sample.
The plotted symbols denote properties of the corresponding star.
Filled circles indicate non-binary Pleiades members.
Open circles indicate non-members observed with \textit{K2}.
Open squares indicate \textit{Kepler} SC stars, and the $\odot$ symbol is the Sun.
Half-filled symbols, or open symbols with diagonal lines, indicate probable
binaries.
All symbols are plotted with colors indicating their $\teff$, as shown on
the color bar.
The dashed line shows the model fitted to all stars except for the 5 excluded
stars (parenthesized on the plot), as described in the text. \label{fig3}}
 \end{figure*} 

\subsection{MFC Amplitude vs Rossby Number}

To understand the physical nature of the MFC amplitude $\sigma_\text{C}$,
we have examined how it varies with the known
global parameters of our stars, namely
($\teff$, $\log(g)$, mass, radius, $P_\text{rot}$), and various combinations of
these.
As one might guess from the behavior
of magnetic indices in stars generally
\citep{noyes1984},
the parameter that predicts observations of
$\sigma_\text{C}$ with the least scatter turns out to be the Rossby number, $Ro \ = \ P_\text{rot}/\tau_\text{c}$,
where $\tau_\text{c}$ (days) is the turnover time for the stellar convection zone,
a timescale that decreases with increasing stellar mass;
for the purposes of this paper we compute $Ro$ from $P_\text{rot}$ and the estimated mass
using Eq.~(11) of \citet{wright2011}.

Figure~\ref{fig3} shows $\sigma_\text{C}$ $vs$ Ro for all stars described in this paper, including
Pleiades members, non-members, likely binaries, \textit{Kepler} short-cadence
stars, and the Sun.
From the Figure, we see that for a range spanning 2 dex at large $Ro$, 
$\log (\sigma_\text{C})$ decreases linearly with increasing $\log (Ro)$. 
For smaller $\log (Ro)$, the amplitude $\sigma_\text{C}$ saturates at what appears
to be a constant value, at least for the range of $Ro$ spanned by this data
set.
The dashed line in Figure~\ref{fig3} represents a simple fit to these data, defined
on a saturated or ``fast'' domain in which
$Ro \leq Ro_\text{sat}$,
and an unsaturated or ``slow'' domain in which
$Ro \geq Ro_\text{sat}$.
We estimated the parameters of this model from a minimum-$\chi^2$ fit
to the observed $\sigma_\text{C}$ values, obtaining

$\log(Ro_\text{sat}) = -1.65$,
with $\log (\sigma_\text{C} )$ modeled by
\begin{equation}
\begin{split}
\text{fast}: & \ \log(\sigma_\text{C}) = 3.59 \\
\text{slow}: & \ \log(\sigma_\text{C}) = 3.59 - 1.24 \log 
\left( \frac{Ro}{Ro_\text{sat}} \right) .
\end{split}
\end{equation}

Almost all of the stars in our sample fall close to the
best-fit $\sigma_\text{C} \ vs \ Ro$ relation.
Cluster non-members follow the same relation as members, except that
the \textit{Kepler} short-cadence stars extend the relation to larger $Ro$
values than are found in the Pleiades.
Likewise, single stars and binaries follow the mean relation about
equally well.
A convenient implication is that our conclusions are unlikely
to be significantly biased by errors in 
binary or membership identification.

We identify 5 outlier stars in the plot that we exclude from the functional fit;
these are flagged in Figure~\ref{fig3} with enclosing parentheses.
Two of these are binaries with fairly small $Ro$ (EPIC211072441 and EPIC211106344),
one a cluster member and one not.
These both have unusual power spectra that are not fit well by the
broken power law model that works well for most stars.
The other excluded stars are 3 of the 4 hottest members of the \textit{Kepler}
short-cadence group of stars, all with $\teff \geq 6300K$.
In section 5.1 we will show that for these stars, leakage of power from
rotating active regions is unusually important.
More details about these and a few other stars with unusual properties are given in Appendix~\ref{appb}.
With those exclusions, the RMS scatter of $\log (\sigma_\text{C})$ around the fit is 
0.22 dex.

The functional relation in Figure~\ref{fig3} is strikingly similar to that
shown by \citet{wright2011} in their Figure~2, 
which displays the ratio of X-ray to bolometric luminosity
as a function of Ro for a sample of  824 Sun-like stars.
These two relations are similar enough that it seems likely that they
arise from analogous processes.
This congruent morphology should not, however, obscure the significant
differences between the two cases.
\citet{wright2011} deals with X-rays from stellar coronal plasma,
whereas $\sigma_\text{C}$ is observed at visible-light wavelengths, and its
color dependence in the solar case identifies it as a photospheric
phenomenon.
Also the X-ray luminosity plotted by \citet{wright2011} represents an energy
loss from the star, while $\sigma_\text{C}$ is the amplitude of a zero-mean fluctuation,
without an obvious connection to the star's energy budget.
In the unsaturated regimes, the two relations have slopes that differ by
about a factor of 2:  for X-rays, $d \log {\rm flux}/d \log (Ro) = -2.7$,
but $d \log (\sigma_\text{C}) / d \log(Ro) = -1.24$.
For $\sigma_\text{C}$, perhaps the analogous signal to consider is not the
RMS fluctuation, but rather the variance.
In this case, the slopes in the unsaturated regimes would be more
nearly equal. 
Still, the most substantial difference between these cases is that the transition between saturated
and unsaturated behavior occurs at different Rossby numbers.
For the X-ray fluxes, \citet{wright2011} find the break between regimes
at $Ro = 0.13$, but the break in $\sigma_\text{C}$ falls at $Ro \ = \ 0.022$,
more than 5 times smaller.\\

 \begin{figure}[t]
 \centering\includegraphics[angle=90,width=\columnwidth]{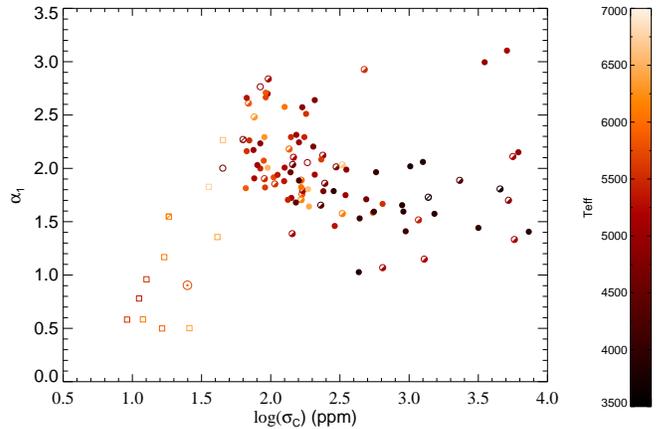} 
 \caption{The power law index $\alpha_1$ of the MFC, shown as a function
of $\log(\sigma_\text{C}$.)
Plot symbols are as described for Figure~\ref{fig3}, and the Sun is shown at lower left. \label{fig4}}
 \end{figure} 

\subsection{Power Law Index $\alpha_1$ vs Rossby Number}

The MFC is characterized not only by its amplitude, but also by its
power law index $\alpha_1$.
Figure~\ref{fig4} shows $\alpha_1$ as a function of $\log (\sigma_\text{C})$.
It suggests that, apart from showing considerable star-to-star 
scatter, $\alpha_1$ displays a rather complex dependence
on $\sigma_\text{C}$.
Plotting $\alpha_1$ against $Ro$ or $P_\text{rot}$ gives diagrams showing behavior
that is similar but not so clear,
especially for small $\sigma_\text{C}$,
corresponding to large $Ro$.
Apart from three apparent outliers, the steepest slopes $\alpha_1$ occur only within approximately $1.6 \leq 
\log \sigma_\text{C} \leq 2.4$,
which from Figure~\ref{fig3} corresponds
roughly to $-0.6 \leq \log Ro \leq -0.1$.
It is not obvious what to make of this dependence, but the behavior is distinctive, so that reproducing
it may prove to be a powerful test of future
physics-based models.
One worrisome feature of Figure~\ref{fig4} is that
the rapid rise in $\alpha_1$ for
$1 \leq \log(\sigma_\text{C}) \leq 1.8$ is
apparent only for the \textit{Kepler} SC stars.
A goal for future work is to measure the properties of the MFC in \textit{Kepler} SC stars with shorter
$P_\text{rot}$, and conversely for \textit{K2} LC stars.

\section{The MFC In Relation to the Global Dynamo}\label{sec5}

Having established some of the properties of the MFC, we now must address more fully
whether we should think of this phenomenon as an artifact of the stellar global dynamo,
or rather as distinct from it.
Is it a process that derives mostly from the existence of a global dynamo,
or does it simply coexist, proceeding (as the granulation does) in parallel 
and governed by its own rules?

The above question is too open-ended to elicit definite answers.
To do better, we consider a few specific ways in which global dynamo
processes might generate effects that resemble the observed MFC.
First, the MFC may be simply numerical leakage of large amplitude,
narrow band rotation
features into the computed power in surrounding frequencies.
This might arise from poorly-constructed window functions, or nonlinear effects
in the time series, or other more complex data processing errors. 
Second, the MFC might arise from diffuse magnetic regions on the target stars
that themselves are born of stellar active regions.
For example, one might imagine long-lived facular regions
being shredded from the boundaries of starspot groups, and eventually becoming
global-scale diffuse magnetic features.
Last, the MFC might come from some small-scale but widespread dynamo process
that requires a seed field (generated by the global dynamo) for its long-term
survival \citep[e.g.,][]{fletcher2010}. 
As we proceed down this list, the line dividing side-effects of
the global dynamo from separate, self-sustaining processes becomes increasingly
unclear.
We might nonetheless gain some insight into the plausibility of scenarios 
such as these by investigating how the amplitude of the MFC behaves
in comparison with various indices of global magnetic activity.

\subsection{Time-Dependent MFC Amplitude vs an Activity Metric for \textit{Kepler} Stars}

The surface manifestations of stellar global magnetic activity vary
erratically with time as active regions form and disperse;
in stars with discernable activity cycles, they also vary more
systematically and on longer timescales. 
To search for connections between activity-based variability
and the strength of the MFC signal in the same star, we subdivided
the time series of photometric variability for each of our sample of 11
stars observed by the \textit{Kepler} mission.
We then broke these time series into nominal ``quarters'', similar but
not identical to \textit{Kepler} quarters --  the differences involve small shifts
in quarter boundaries to place data gaps between the new quarters,
rather than within them.
Partitioned in this way, the time series for all stars have between 11 and
13 valid quarters, and the quarters have average durations of 84 days.
Given these partitions, we computed the $PSD(\nu)$ for each star $i$ and each nominal quarter
$j$, and integrated appropriately over frequency to yield quarter-dependent 
harmonic and
mid-frequency continuum indices $\sigma_{\text{H}ij}$ and 
$\sigma_{\text{C}ij}$.   

\begin{figure}[t]
 \centering\includegraphics[angle=90,width=\columnwidth]{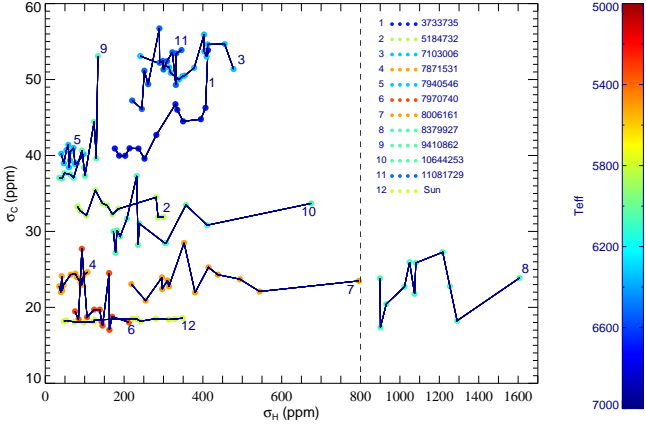} 
 \caption{$\sigma_\text{C}$ plotted against $\sigma_\text{H}$
for the 11 \textit{Kepler} short-cadence stars in our sample, and the Sun.
Each star is represented by a series of connected dots, corresponding to
the RMS values measured by \textit{Kepler} for each of between 9 and 13 ``quarters'',
as explained in the text.
Plotted points are color-coded according to $\teff$.
Note that for $\sigma_\text{H}$ values above 800, rightward of the vertical dashed line, the horizontal scale is compressed by a factor of 2. \label{fig5}}
 \end{figure}

Figure~\ref{fig5} shows the measured $\sigma_{\text{C}ij}$ as a function of
$\sigma_{\text{H}ij}$, plotted for all stars $i$ and quarters $j$.
The quarters for each star are plotted in order of increasing
$\sigma_{\text{H}ij}$, and connected with solid lines.
Our aim in this Figure is to display explicitly any functional connection
between $\sigma_\text{H}$ and $\sigma_\text{C}$.

The stars in this sample show $\sigma_\text{H}$ variability across quarters
that typically spans a factor of 4,
whereas the variability range in $\sigma_\text{C}$ is typically only 20\%.
Further, most of the stars in Figure~\ref{fig5} show little or no 
correlation between changes
in $\sigma_\text{H}$ and in $\sigma_\text{C}$.

This lack of correlation is good evidence that
the processes that we measure with $\sigma_\text{C}$ and $\sigma_\text{H}$ are,
in most stars, physically distinct.
For the same reason, we conclude that numerical
power leakage between the frequency regimes
below and above $\nu_0$ is not important.
For the hot stars where significant correlation
is present, we think it most likely that
basic differences in the structure of near-surface convection
have shortened the timescales associated with
global magnetic activity,
moving power from low $\nu$ into the $\nu$ range that we
identify with $\sigma_\text{C}$.
How this might happen is an interesting question,
but one that is beyond the scope of this paper.

\subsection{Inclination Angle}

One expects variability of rotating stars to depend to some degree on 
the inclination $i$ of the rotation axis to the
line of sight.
This may occur for two different reasons.
First, a portion of the a star's photometric variability arises from long-lived brightness structures being carried
across the stellar limb by rotation.
At small inclination, these transitions slow down, causing the power spectrum of their variability to
compress towards smaller temporal frequencies.
In the extreme pole-on case ($i=0$), one sees no rotation-driven variability at all, and brightness
fluctuations result only from the intrinsic time evolution of the features.
Second, when rotation is fast enough to affect the internal dynamics of brightness features,
their morphology and time evolution
may depend on the stellar latitude.
In this case, the observed brightness power spectrum
may depend on whether one is observing mostly polar
regions (for small $i$), or equatorial ones
(for $i \simeq \pi /2$).

We have therefore searched for inclination-dependent variation of the observed power spectra.
This is feasible because, for the bulk or our observed
sample, there are observations in R16 of both $P_{\text{rot}}$ and the projected rotational
speed $v \sin i$.
We thus estimate for each star $\sin i = v \sin i/v_{\text{Eq}}$,
where $v_{\text{Eq}}$ is the estimated equatorial rotation speed $v_{\text{Eq}}=2\pi R_*/P_{\text{rot}}$, and $R_*$ is the stellar radius.
We have data to compute this estimate for 77 of our sample stars, all but one of these being Pleiades
cluster members.

The inclination values we derived are distributed as expected for a sample that has
randomly oriented rotation axes:
about half of the sample shows $\sin i \geq 0.8$,
while only 6 stars have $\sin i \leq 0.5$, and 3 of these
have $\sin i \leq 0.4$;
because of errors in the estimates of $v \sin i$ and $v_{\text{Eq}}$, about a quarter have $\sin i > 1$.

The results of comparing $\sin i$ with our various photometric indices ($\sigma_{\text{H}}$, $\sigma_{\text{C}}$, $\alpha_1$, as well as the
measures of integrated power and spectrum slope
for frequencies below $\nu_0$)
are, however, entirely inconclusive:
none of these parameters show a credible dependence on $\sin i$.
We have also examined the power spectra of the 6
sample stars with $\sin i \leq 0.5$,
and compared them with the spectra of other stars
having similar $P_{\text{rot}}$, finding no obvious
differences.

We conclude that evidence suggests that the
photometric fluctuations at frequencies above $\nu_0$
arise mostly from the intrinsic time evolution of
brightness features, rather than from rotational modulation of small, essentially static features.
Quantifying this conclusion in a meaningful way will likely require significantly better estimates of
$v \sin i$ and of stellar radii.

\subsection{Dependence on the Rossby Number}

If the MFC is merely an artifact of the global dynamo, then we would expect
the observables $\sigma_\text{C}$ and $\sigma_\text{H}$ to vary 
in similar ways with variations in controlling parameters
(notably the Rossby number).
Real stellar behavior may confound this expectation if one or both observables
are unreliable indicators of some more fundamental internal
state of the star.
Nevertheless, it is informative to ask if $\sigma_\text{C}$ and $\sigma_\text{H}$ respond in
similar ways to changes in $Ro$.
The answer to this question is implicit in Figures~\ref{fig3} and \ref{fig5}:  $\sigma_\text{C}$
is a fairly tight function of $Ro$, but $\sigma_\text{H}$ is poorly correlated
with $\sigma_\text{C}$, so $\sigma_\text{H}$ must not depend on $Ro$ in the same way
that $\sigma_\text{C}$ does.

\begin{figure}[t]
 \centering\includegraphics[angle=90,width=\columnwidth]{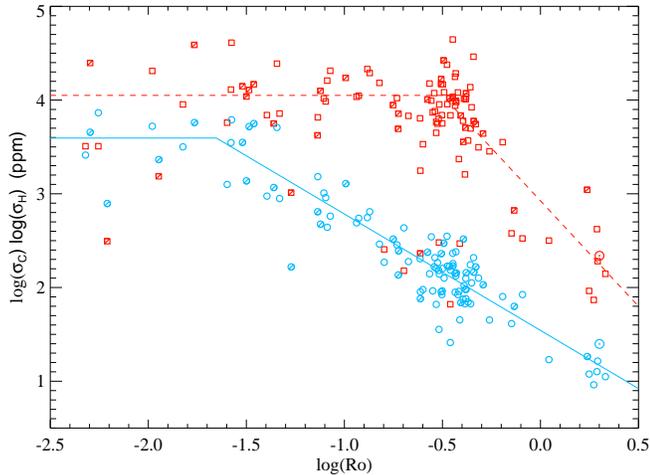} 
 \caption{$\log (\sigma_\text{C} )$ (blue circles) and $\log (\sigma_\text{H} )$
(red squares) for the entire sample of stars, shown as a function
of the log of the Rossby number $\log (Ro)$.
Plot symbols with diagonal bars indicate probable binaries.
Also plotted are simple fitted models of the dependence on Ro of
the two quantities.\label{fig6}}
 \end{figure}

Figure~\ref{fig6} makes this inference explicit, by showing both $\sigma_\text{H}$ and
$\sigma_\text{C}$ $vs$ $Ro$, for the entire sample of stars. 
Evidently $\sigma_\text{C}$ is strongly dependent on $Ro$, whereas $\sigma_\text{H}$
is roughly independent of it, except perhaps near
the upper limit of $Ro$ represented here.

Indeed, from Figure~\ref{fig6} one sees that the relation $\sigma_\text{H} (Ro)$ follows the same generic form
as does $\sigma_\text{C} (Ro)$, except that the saturated regime extends to
$\log (Ro) \approx -0.5$, and the scatter
of $\sigma_\text{H}$ is much larger than for
$\sigma_\text{C}$.
Also, note that the stars that fall far
(sometimes 2 dex) below the upper envelope
of $\sigma_\text{H}$ are not preferentially
binaries.
Thus, $\sigma_\text{H}$ may be saturated for almost all the stars reported here,
with only the very slowest-rotating stars occupying the unsaturated regime.
Other work \citep{vansaders2016, metcalfe2017, brandenburg2017}
suggests that stellar dynamos show qualitatively different behavior
at long rotation periods, starting at $Ro$
near the solar value.

\section{Discussion}\label{sec6}

Up to this point, our discussion of the MFC phenomenon has been almost
entirely empirical and descriptive.
We now turn to a more physics-based discussion,
necessarily conjectural because of our incomplete
knowledge of the processes that create the MFC.
We will proceed in three steps.
First we describe a few relatively assumption-free
inferences that we may derive from the observations
described above.
Next, we attempt to relate the MFC as observed in
the Sun to known solar processes.
Last, we consider how the surface structures of
rapidly-rotating stars might differ from those seen
on the Sun, in order to produce the large
observed values of $\sigma_\text{C}$ that are seen in
rapid rotators.

\subsection{Physical Inferences}

From the Blue/Green/Red color dependence of the solar $\sigma_\text{C}$ as observed
by VIRGO/SPM, the process generating MFC photometric variability must be
almost entirely thermal, with a temperature corresponding roughly to that
of the stellar photosphere.

On the other hand, the large variation in $\sigma_\text{C}$ across stars with similar
structural parameters (mass, age, composition, $\teff$) 
suggests, by the absence of alternative explanations, that the underlying process is magnetic.
This idea is reinforced by the tight relation between the MFC
amplitude $\sigma_\text{C}$ and rotation, specifically the Rossby number $Ro$.
Since $Ro$ is a measure of a convection zone's ability to generate magnetic
helicity, 
and since it is connected to other magnetic processes in stars
\citep[e.g.][]{noyes1984, wright2011}, it appears likely that the MFC is connected to some process
involving a magnetic dynamo.

The temporal power spectrum of the MFC is a continuum, by definition lacking notable 
modulation of the photometric fluctuations by the stellar rotation. 
This implies that whatever process generates the MFC signal, it must be
distributed almost uniformly in stellar longitude, and possibly uniformly 
over the entire surface of the star. 
Therefore, large-amplitude signals from a few localized spots or active
regions do not yield a viable model of the observed fluctuations.

If we suppose that the MFC process presents a cellular morphology, 
like granulation, present almost
everywhere on the stellar surface all the time, then we can place rough limits
on the properties of the cells.

Thus, from the low-frequency limit of the MFC ($\nu_0 \approx 20 \mu$Hz), 
we infer a cell lifetime $\tau_\text{cell}$ of about $1/\nu_0$, 
or roughly half a day.
If we equate $\tau_\text{cell}$ to the cell turnover time, and assume that flow
speeds $v_\text{cell}$ within the cell are no larger than typical 
macroturbulent velocities
of about 3 km/s, then we estimate that
$2 \pi r_\text{cell} / v_\text{cell} \leq \tau_\text{cell}$,
so $r_\text{cell} \leq$ $1.6 \times 10^4$ km.
On the Sun, this is within a factor of 1.5 of the scale of solar supergranules \citep{rincon2018}.

Assuming that the MFC results from the hemispheric average of independent cellular structures,  
the net photometric
signal $\sigma_\text{C}$ from a star should be given by
$\sigma_\text{C} = \sigma_\text{cell} / \sqrt{N_\text{cell}}$,
where $\sigma_\text{cell}$ is the RMS relative 
brightness variation of one cell, and
$N_\text{cell}$ is the number of distinct
cells in a hemisphere.
Thus, to achieve a large value of $\sigma_\text{C}$,
we require $\sigma_\text{cell}$ to be large or
$N_\text{cell}$ to be small, or both. 

To summarize: observations suggest that the Mid-Frequency Continuum
(MFC) observed in the photometric fluctuation spectra of Sun-like stars
is associated with a photospheric process that has a significant magnetic
component, and a timescale similar to that of solar supergranulation.
Constraints on the dominant spatial scale of the process are weak, 
but are consistent with a ubiquitous cellular spatial structure with 
typical cell sizes similar to those of solar supergranules. 

\subsection{Back to the Sun}

We would like to establish a tentative connection between
the structures generating the MFC and known phenomena on the Sun.
To conform with our estimated MFC lifetimes and length scales,
to satisfy the requirement of near-uniform presence on the whole star surface,
and to place the main source of radiation in the photosphere, 
we are led to consider supergranulation, and features related to it.

Since the supergranulation phenomenon is not
commonly encountered outside of solar physics,
we draw here on the recent review by \citet{rincon2018} to summarize its relevant properties.
Supergranulation is the largest cellular circulation pattern visible at the
solar surface.
Many of the properties of supergranulation are well known, although
controversies remain about its underlying nature.
For instance, there is little consensus about what sets the size scale of
supergranules, or even whether one should think of them as flows driven
by thermal convection.
Solar supergranules have typical diameters of about 35 Mm,
or about 5\% of the solar radius.
Estimates of their lifetimes vary, depending on definitions, but the most
commonly cited values are between 1 and 2 days.
Because of their large horizontal scale 
relative to the photospheric scale height,
their velocity fields are mostly horizontal, with flows diverging
from the cell centers at typically 300-400 m/s.

Supergranular flows sweep magnetic fields with them to the cell boundaries,
where they become entrained in downflows that compress the fields
into small (100-500 km) flux tubes with field strengths of one to a few kG.
Collectively, these tubes make up the magnetic network \citep{bellot_rubio2019}.
Flux tubes often appear in broadband emission as
photospheric bright points -- the result of lowered opacity in the partially
evacuated tubes allowing radiation from deep, hot layers of the solar 
atmosphere to escape \citep{spruit1976,libbrecht1984}.
If one excludes
strongly magnetized regions
(the network downdrafts), then supergranules
display only very little brightness structure (with
typical contrast of $7 \times 10^{-4}$, \citet{goldbaum2009})
Including the magnetized regions, however, supergranules show broad-band
relative flux variations with spatial RMS of 2 to 3$\times 10^{-3}$,
depending on wavelength \citep{lin1992}.

We wish to
estimate the supergranular contribution to the temporal RMS of the disk-integrated solar flux,
so we can compare this to the VIRGO/SPM measurements described above.
To do this, we assume that \citet{lin1992}'s numbers for the spatial RMS of the supergranular contrast
also describe the temporal RMS,
and we take the number of supergranules per hemisphere to be $N_\text{cell} = 2500.$
Then we extrapolate the \citet{lin1992} red and green measurements to the VIRGO/SPM blue wavelength,
assuming that these scale according to the temperature
derivative of the Planck function.
This process gives estimated supergranule RMS for
\{blue, green, red\} colors equal to \{60, 40, 25\} ppm.
These agree within better than a factor of 2 with the VIRGO/SPM measurements of $\sigma_\text{C}$, namely \{blue, green, red\} = \{46.0, 23.4, 13.7\} ppm.

We conclude that the VIRGO/SPM observations are broadly consistent with the idea
that the solar MFC arises from supergranulation and its associated magnetic
phenomena, in particular photospheric bright points.
We warn, however, that this apparent consistency is by no means a proof of identity.
Such a proof will require both a reliable quantitative theory of supergranulation and the
magnetic network,
and also more comprehensive and directed observations.

These uncertainties notwithstanding, we now hypothesize that the MFC observed in the Sun's disk-integrated
flux arises from photospheric bright points located in the quiet-Sun magnetic
network that delineates supergranule boundaries.
Individual bright points have lifetimes of
tens of minutes to a few hours \citep{giannattasio2018},
but the network structures to which they belong
may survive for considerably longer.
We then suggest that it is the formation and decay of these tubes (with their corresponding radiative
signatures) that leads to the power-law temporal spectrum defining the MFC.
The timescales and inferred spatial scales of the MFC then
arise partly from the lifetimes and peculiar motions
of the individual bright points, but
mostly from the lifetimes and sizes of the supergranules' organizing flows.
More work is needed, of course, to see if this picture is accurate.

\subsection{Extension to More Active Stars}

If the Sun's MFC results from photospheric bright points, then what of
the MFC seen in rapidly-rotating Pleiades members, in the saturated regime
of Figure~\ref{fig3}?
If we adopt the idea that the MFC results from a hemispheric average of
signals from uncorrelated cellular structures,
then attaining the typical saturated photometric $RMS$ of 0.4\% means that
$\sigma_\text{C}=\sigma_\text{cell} / \sqrt{N_\text{cell)}}$ for such stars must be larger
than the solar $\sigma_\text{C}$ by a factor of about 50.

It is plausible that both $\sigma_\text{cell}$ and
the size or shape of overturning cells (hence $N_\text{cell}$)
might depend on the Rossby number.
Thus, in dynamos, $Ro$ governs the creation rate of magnetic helicity;
in non-magnetic convection it measures the relative strength of Coriolis
and buoyancy forces.
But our present knowledge, both observational and theoretical, is not
sufficient to go much beyond these simple statements.
For now, we restrict ourselves to four comments.

First, since the filling factor of bright points
in the quiet Sun is roughly 1\% \citep{giannattasio2018}, it seems implausible
that fast rotation (or any other process)
can increase their number density on the stellar surface by the required factor of 50.
Put differently, if we ascribe all of the $Ro$-related variation in $\sigma_\text{C}$ to
variations in $\sigma_\text{cell}$,
then the required $\sigma_\text{cell}$ at saturation is about 10\%.
This is comparable to the contrast seen in solar granulation, but on
spatial and temporal scales that are about an order of magnitude larger.
It is not obvious how the fields and flows might organize themselves to
achieve this.
We therefore speculate that as $Ro$ decreases, both $\sigma_\text{cell}$ and $N_\text{cell}$
are involved in the growth of $\sigma_\text{C}$.
That is, the flows and fields
responsible for the MFC not only become more intense, but likely also
undergo significant changes in the size and shape of their organizing cells.
Alternatively, with the stronger non-global fields at low Ro, the fields may start to interfere with the convective
energy flow from the interior and produce short-lived pores and small
starspots.  
This could produce the observed photometric variability
without involving as much surface area as would bright points, because
spot brightness contrasts are larger. 

Second, in the solar literature one finds discussion as to whether the supergranular 
intranetwork magnetic fields result from a
near-surface magnetic dynamo that operates
independently from the global one that is 
responsible for sunspots and the solar magnetic cycle \citep{vogler2007, lites2011, bellot_rubio2019}.
We observe (e.g., in Figure~\ref{fig5}) 
a general lack of correlation, pronounced in
the solar case, between
the strength $\sigma_\text{C}$ of the MFC phenomenon and $\sigma_\text{H}$, the corresponding index of global magnetic activity.
This poor correlation is evidence for two or more different dynamos.

Third, Fig. 3 suggests there is a small-scale dynamo that at small $Ro$ grows in strength to compete with
the global one,
or perhaps that the rotation-dominated global dynamo
extends to smaller spatial scales at low $Ro$.
In either case, for rapidly-rotating stars one should expect to see stronger fields at small spatial
scales, hence increased magnetic complexity,
and hence weakened braking torque exerted by the stellar wind \citep{garraffo2015, garraffo2016, garraffo2018, reville2015}.
Such a torque reduction may provide a mechanism to explain the bimodal $P_\text{rot}$ distributions
seen in young star clusters such as the Pleiades \citep{soderblom1993, hartman2010, barnes2010, brown2014}.

Last, we find it curious that the relation between $Ro$ and $\sigma_\text{C}$,
shown in Figure~\ref{fig3}, shows so little scatter.
The MFC range of timescales suggest turnover times and cell depths that are considerably smaller than those for the observed stars' entire convection zones. 
But  naively, we expect that the relevant Rossby number for these flows should relate to the
turnover times for the shallow flows,
while the Rossby numbers used in Figure~\ref{fig3} are computed using $\tau_\text{C}$, the turnover time for the entire convection zone.
Thus, to get the fairly tight relation between
$\sigma_\text{C}$ and $Ro$ that is shown in
Figure~\ref{fig3}, it must be that the MFC flow
turnover times scale with $\teff$ in the same
way as does $\tau_\text{C}$.
It is not obvious why this should be so.
Moreover, at least three other stellar magnetic phenomena show amplitude
$vs$ $Ro$ relationships that resemble the one we plot in Figure~\ref{fig3}.
These are the harmonic amplitude $\sigma_\text{H}$ shown in Figure~\ref{fig6}, the coronal X-ray emission
described by \citep{wright2011}, and the unsigned magnetic flux
measured by \citet{see2019}.
Some of these processes appear to be physically
related, in that the photospheric magnetic network,
the chromospheric network, and the footpoints of X-ray
loops are often co-spatial.
This makes sense if the different heights at which these phenomena occur are linked by strong, more-or-less vertical magnetic fields.
But if these various phenomena reflect a single underlying dynamo process, then
why are their governing parameters (break point in $Ro$, slope of unsaturated
regime) so different from one to another?
If they represent physically different dynamos, why are their
morphologies so similar?
Also,why is the scatter in these relationships so much smaller for $\sigma_\text{C}$
and for the X-ray flux than for the other two cases?
To answer these and related questions we will likely need better
observational statistics,
new kinds of observational diagnostics, and improved quantitative
models of magnetized flows in stellar atmospheres.

We expect that the Mid-Frequency Continuum will prove important in understanding stellar rotation and magnetism, both because MFC properties are relatively easy to observe,
and because, once understood, the MFC may well provide powerful diagnostic
tools with which to probe stellar convective dynamics and magnetic activity.
With respect to observability, note that
the MFC observables $\sigma_\text{C}$, $\alpha_1$ 
may be recovered from
data sets of only a few days duration, needing a sampling cadence of only
one or two samples per hour, and having (by the standards that 
apply to space photometry) relatively poor photometric precision.
For example, for almost the whole sky down to $V$ magnitudes of
12 or so, the TESS mission \citep{ricker2014} provides 27-day-long near-continuous
stellar time series with a 2-minute sampling cadence (for targeted stars) and $S/N$ adequate
for this purpose.
Thus, MFC parameters may be obtained for vast samples of
stars.
As for diagnostic power, the dependencies shown in Figures~\ref{fig3} and \ref{fig4} are fairly well defined,
and can be improved with larger samples of stars. 
Also they are complex enough that the 
ability to reproduce them, even in part, would provide a strong
validation of future numerical models.\\

We dedicate this paper to the memory of John Stauffer, outstanding human being, good friend, and premier source of knowledge about the Pleiades.
The authors would like to thank Jamie Tayar for assistance obtaining stellar parameters from APOGEE DR16.
We thank Matthias Rempel and Mark Rast for enlightening conversations,
and also the anonymous referee, whose careful reading and thoughtful suggestions
improved this paper in many ways.
This paper includes data collected by the \textit{Kepler} mission and obtained from the MAST data archive at the Space Telescope Science Institute (STScI). Funding for the \textit{Kepler} mission is provided by the NASA Science Mission Directorate. STScI is operated by the Association of Universities for Research in Astronomy, Inc., under NASA contract NAS 5–26555.
This work has made use of data from the European Space Agency (ESA) mission
{\it Gaia} (\url{https://www.cosmos.esa.int/gaia}), processed by the {\it Gaia}
Data Processing and Analysis Consortium (DPAC,
\url{https://www.cosmos.esa.int/web/gaia/dpac/consortium}). Funding for the DPAC
has been provided by national institutions, in particular the institutions
participating in the {\it Gaia} Multilateral Agreement.
In this work we made use of 
the APOGEE database \citep{ahumada2020}, 
and the EPIC star catalog \citep{huber2016}.
This research has made use of the WEBDA database, operated at the Department of Theoretical Physics and Astrophysics of the Masaryk University \citep{paunzen2008}.
R.A.G.\ acknowledges the support from PLATO and GOLF CNES grants. S.M.\ acknowledges support by the Spanish Ministry of Science and Innovation with the Ramon y Cajal fellowship number RYC-2015-17697 and the grant number PID2019-107187GB-I00. T.S.M.\ acknowledges support from NASA grant 80NSSC20K0458.  A.R.G.S.\ acknowledges support from NASA grant NNX17AF27G and STFC consolidated grant ST/T000252/1.

\appendix
\section{DATA REDUCTION AND FITTING PROCEDURES}\label{appa}

For the Pleiades long-cadence (LC) time series, our data analysis proceeded through
4 steps:
(1) pre-processing, (2) power spectrum computation, (3) fitting the
coherent (harmonic) part of the spectrum, (4) fitting the continuous spectrum.

\subsection{Preprocessing}
The EVEREST time series from which we worked already consisted of weighted sums
of \textit{Kepler} pixels, corrected for various instrumental effects as described
by \citet{luger2016}.
We did several comparisons of the EVEREST time series with other
methods [VARCAT \citep{2016MNRAS.456.2260A} and \textit{K2}SFF \citep{2014PASP..126..948V}]. 
For the latter two time series we followed the same pre-processing as for the EVEREST and \textit{Kepler} light curves as described in Sect.~\ref{Sect3.1}. It was found
that the EVEREST data best served our purposes, because they gave the highest $S/N$ ratio near peaks in the power spectrum.

The inputs to our Pleiades pipeline were zero-mean,
high-pass filtered and apodized time series,
sampled once per 29.42 minutes,
in units of parts-per-million (ppm) of the mean stellar brightness.

\subsection{Power Spectrum}

For LC data, we used the \citet{scargle1982} algorithm to compute the power
spectrum on a grid with resolution 0.1 $\mu$Hz.
Given the duration of the LC time series, this oversampled the true resolving
power of the data strings by a factor of about 1.3.
For the SC data, we computed the power directly from the
discrete Fourier transform of the time series.
The resulting resolution was about 0.011 $\mu$Hz.
We normalized the power spectrum to give the power spectral
density ($PSD$), in units of ppm$^2/ \mu$Hz.

It is computationally convenient to frame the spectrum fitting as
a least-squares problem. 
But to avoid biased results, the data being fitted should
have errors that are more or less normally distributed, unlike
the ($\chi^2$ with 2 degrees of freedom) distribution that emerges from
the power spectrum computation.
To get unbiased fits, we therefore smoothed the spectra in frequency, using
pseudo-Gaussian kernels with widths of about 5 frequency elements
(for LC spectra) or 19 elements (for SC spectra).
This smoothing is acceptable, because
in fitting the continuum power, we do not require even the full frequency
resolution of the LC data.

The fitting procedure we used is a sequential one
that first fits the coherent (line) part of the spectrum, and then uses the
residuals from this fit to model the continuous part.
In the future we hope to develop a more elaborate fitting code, which will
fit the entire spectrum simultaneously and consistently, and moreover will
be completely automated, removing the biases and lack of repeatability that
go with human intervention.
Such a code would also make it practical to analyze many more stars than we
are now capable of doing.
For the present, however, we find it most expedient to proceed with the
current analysis tools, bearing their shortcomings in mind.

\subsection{Fitting Rotational Harmonic Components}

Rotating spotted stars produce observed photometric
time series that are often dominated by the star's rotation frequency and
its harmonics.
Accounting for these multiple large-amplitude narrow-band signals is the
most complicated part of our spectrum-fitting process, but for our
immediate purposes (fitting the continuum spectrum) its results are largely irrelevant.

We represent the coherent part of each power spectrum as 
\begin{equation}
 P_\text{harm}(\nu) = \sum_{i=1}^{N_s} \left ( \sum_{j=1}^{N_h} A_{ij} W_i (\nu -   
      j \nu_i) \right ) \ .   
\end{equation}
Here the sum over harmonic sequences $i$ is itself a sum of terms 
related to different stellar
rotation frequencies (e.g., the various members of a multiple star system,
each with its own rotation frequency, or different latitudes on a single
differentially-rotating star).
For most stars, this sum over distinct rotating systems involves only
a single system.
Each of the $N_s$ rotating systems is taken to have its own 
line profile $W_i$, which
we take to be shared by all of the associated rotation harmonics, having
frequencies $\nu = j \nu_i$ for $j$ = {1,2,...N$_i$}.
Finally, the various harmonics $j$ composing the rotating system $i$ have
amplitudes $A_{ij}$, each taken to be independent of all the others.
The parameters required for a fit are the number of rotation systems $N_s$,
and for each of these the corresponding fundamental frequency $\nu_i$,
the parameters of the line profile $W_i$, the number of harmonics $N_i$,
and the amplitudes of all the harmonics $A_{ij}$.

We parameterize the shape of the line profile $W_i$ as a generalized
Lorentzian
\begin{equation}
W_i(d \nu ) = 1 / (1 + | d \nu / s|^m ) \ ,
\end{equation}
where $s$ is a width parameter and $m$ is a power-law index, which would
be 2 for a true Lorentzian.
The code takes $m = 4.5$ for all lines, but fits $s$ independently for each
rotation system, fitting to the observed profile of the fundamental frequency
$\nu_i$ for the system.
(This is usually, though not always, the largest peak in the spectrum.)

We fit the model parameters to the smoothed observed power spectrum via
$\chi^2$ minimization, assuming uncertainties $\sigma = PSD/N_{sm}^{1/2}$,
where $N_{sm}$ is the effective smoothing width.

 \begin{figure}[t]
 \centering\includegraphics[angle=90,width=5.5in]{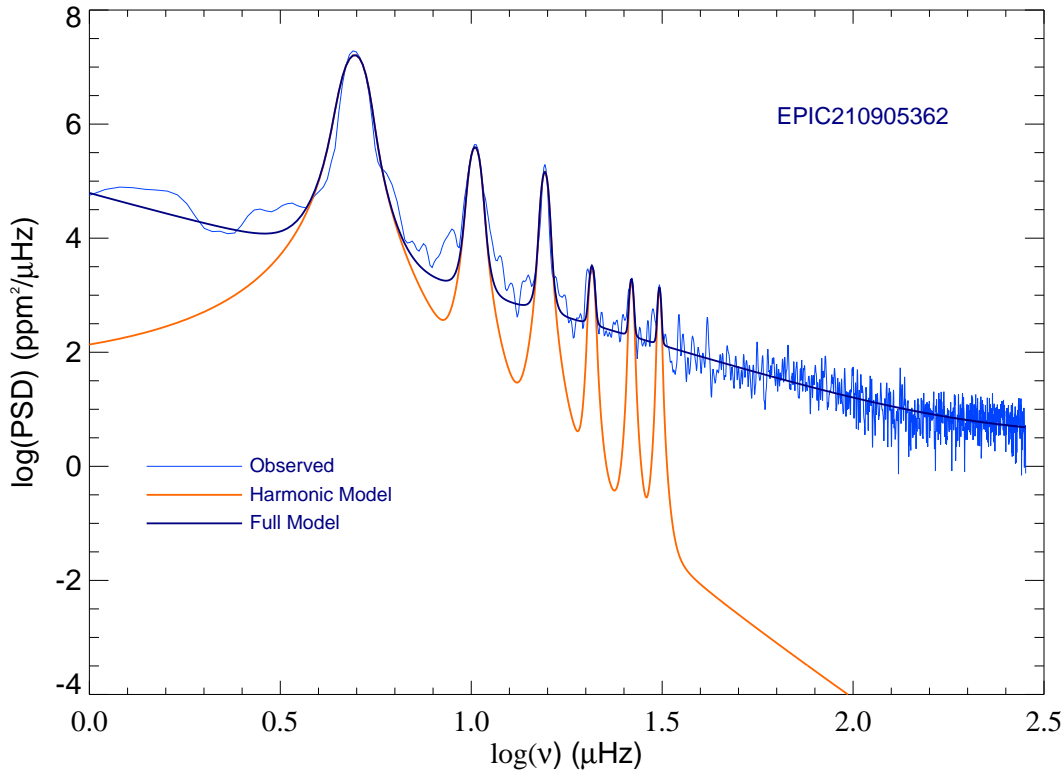} 
 \caption{Power spectral density ($PSD$) of brightness fluctuations
of a typical Pleiades member, having
$P_\text{rot} = 2.379$ days.
The thin blue trace is the observed spectrum.
The thick orange trace is the fitted model of the harmonic part of the power spectrum.
The thick black trace is the full model, $i.e.$,
the sum of the harmonic and continuum models.\label{fig7}}
 \end{figure} 

Figure~\ref{fig7} shows in orange the result of fitting the coherent part of the power
spectrum of a typical star (EPIC210905362) in the way just described.
Evidently the fit succeeds in capturing the broad features of the coherent
spectrum, in particular the total power integrated across each line profile.
The fit does not, however,  accurately reproduce the line shapes outside the
line core regions, leaving substantial residuals in the near wings of the lines.

\subsection{Continuum Fitting}

We proceed in the continuum-fitting step simply by 
subtracting the fitted coherent spectra as best we can, and assigning small 
or zero fitting weights to the residuals  within 7 line widths of the 
modeled line center frequencies.

After subtracting the coherent signal from rotating active regions, we model
the remaining power spectra over the frequency range 
1 $\leq \nu \leq$ 285 $\mu$Hz as a continuous piecewise power law, using
only two frequency domains, with a break point at the frequency $\nu_0$:
\begin{equation}
\begin{split}
\nu \geq \nu_0 :\ & P_\text{cont} = A \left ( \frac{\nu_0}{\nu} \right )^{\alpha 1}\\
\nu \leq \nu_0 :\ & P_\text{cont} = A \left ( \frac{\nu}{\nu_0} \right )^{\alpha 2}.  
\end{split}
\end{equation}
By performing the fit in $\log (\nu)$, $\log (PSD)$ coordinates, this reduces to
fitting for the $P_\text{cont}$ amplitude $A$, and the two straight line slopes 
$\alpha_1$, $\alpha_2$.
We perform this fit via $\chi^2$ minimization, using as input the $\log$ of
the $PSD$ residuals from the coherent harmonic spectrum fit described above.
We compute weights based on the smoothed $\chi^2$ distribution also described
above. 
Figure~\ref{fig7} shows in black the result of such a fit applied to the same star,
using $\nu_0 = 20 \mu$Hz.
This is a typical case, giving a fit that represents the observed spectrum
in a way that is satisfactory, though not perfect.

In the fits described in the main text, we use
$\nu_0 = 20 \mu$Hz for fitting all stars.
Since we do not yet understand the physics involved, there is no $a \ priori$
reason to believe that $\nu_0$ should be the same for all stars.
But neither do we have justification for any particular dependence on
stellar parameters.
So for the current study  we adopt $\nu_0 = 20 \mu$Hz for all stars, 
justified by simplicity and by
the fact that almost always this value delivers sensible fits.

\section{EXCEPTIONAL STARS}\label{appb}

A few stars are notable exceptions to
the systematic behaviors that characterize most stars.
In the upper left corner of Figure~\ref{fig5} one finds 3 stars 
(KIC~11081729, KIC~7103006, KIC~3733735) that stand out for large values of
$\sigma_\text{C}$, and for relatively strong correlation between $\sigma_\text{C}$ and
$\sigma_\text{H}$.
As the color coding suggests, these are the hottest stars in our \textit{Kepler} SC
sample;  all have $\teff \geq$ 6300 K.
Only 5 stars in our Pleiades sample have such large $\teff$.
As well, these stars have relatively short rotation periods, and they fall
about 0.5 dex below the mean $\log Ro$-$\log \sigma_\text{C}$ relation.
All of these characteristics suggest that these 3 stars lie near the boundary
of the Kraft break \citep{kraft1967}, having shallow and 
inefficient convection zones that
cause them to be unrepresentative of our larger star sample.
Also anomalous is KIC~8379927, with $\sigma_\text{H}$ both unusually large and
unusually variable.
This star is one of the LEGACY sample of stars with high-quality \textit{Kepler}
asteroseismic measurements \citep{lund2017}.
Its time series shows what appear to be beats between fundamental rotation
harmonics separated by about 0.05 $\mu$Hz.
Speckle \citep{horch2012} and radial velocity \citep{griffin2007} 
observations show this 
star to be binary, with a companion that is
about 1.4 magnitudes fainter than the primary.
It is not clear whether its peculiar activity parameters arise from its fairly 
bright companion, from its large apparent differential rotation, 
or from some other cause.

\clearpage

\startlongtable
\movetabledown=50mm
\begin{rotatetable}
\begin{deluxetable}{rrrrrrrrrrrrrrrrr}\label{tab1}
\tablecolumns{17}
\tabletypesize{\footnotesize}
\tablecaption{Star sample physical and fitted parameters}
\tablehead{
\colhead{Starname} & \colhead{Hz} &
\colhead{RA} & \colhead{Dec} &
\colhead{$V$} & \colhead{$V-K$} &
\colhead{Mass} & \colhead{$T_{\rm eff}$} & \colhead{$\log(g)$} &
\colhead{$P_{\rm rot}$} &
\colhead{$Ro$} &
\colhead{Bin}&
\colhead{Memb} &
\colhead{Cdnc} &
\colhead{$\log\sigma_\text{H}$} &
\colhead{$\log\sigma_\text{C}$} & \colhead{$\alpha_1$}
}
\startdata
EPIC211153286&  8508&  58.57837&  25.49526& 12.87&   3.07&  0.73&  4226& 4.76&  7.74&  0.312&  b&   nm& LC&   4.22&   2.36&   1.655\\
EPIC211149600&  8544&  57.64885&  25.42650& 11.96&   2.38&  0.85&  5018& 4.57&  0.31&  0.017& pb&   ok& LC&   4.59&   3.76&   1.333\\
EPIC211147822&  8545&  57.66375&  25.39375& 12.33&   2.89&  0.87&  4763& 3.42&  0.58&  0.033&  b&   nm& LC&   4.11&   3.72&   1.701\\
EPIC211129400&      &  56.42688&  25.05709& 13.60&   3.09&  0.68&  4542& 4.67&  2.84&  0.115&  s& best& LC&   4.04&   2.69&   1.710\\
EPIC211129308&      &  57.75964&  25.05546& 13.37&   2.99&  0.82&  4903& 4.58&  5.79&  0.299&  s& best& LC&   3.75&   2.26&   2.054\\
EPIC211121141&  0451&  56.20907&  24.91113& 13.57&   3.28&  0.71&  4642& 4.64&  5.68&  0.243&  s& best& LC&   3.81&   2.31&   2.204\\
EPIC211118542&  0885&  56.53240&  24.86687& 12.17&   2.81&  0.90&  5043& 3.11&  6.95&  0.414&  b&   nm& LC&   3.71&   1.98&   2.839\\
EPIC211117077&  0191&  55.96730&  24.84162& 14.42&   3.82&  0.51&  3950& 4.84&  3.00&  0.085&  s& best& LC&   4.31&   2.76&   1.964\\
EPIC211114317&  0314&  56.08369&  24.79618& 10.43&   1.52&  1.10&  5574& 3.89&  1.47&  0.118&  s& best& LC&   4.05&   2.74&   1.585\\
EPIC211113345&      &  56.05799&  24.77939& 10.84&   1.77&  1.04&  5814& 4.44&  4.03&  0.294&  s& best& LC&   3.65&   1.82&   1.814\\
\enddata
\tablecomments{Table~\ref{tab1} is published in its entirety in the
electronic edition of the {\it Astrophysical Journal}.  A portion is
shown here for guidance regarding its form and content.
Columns:  Starname; RA, Dec = Right Ascension, Declination (degrees);
$Hz$ = Hertzsprung Catalog number from WEBDA \citep{paunzen2008};
$V$ = Johnson V magnitude; 
$V-K$ = $V-K$ Color Index (magnitudes); 
Mass = Stellar Mass ($M_\sun$); 
$\teff$= Effective Temperature (K); 
$\log(g)$ = $\log$ Surface Gravity (cgs); $P_\text{rot}$ = Rotation Period (days); $Ro$ = Rossby Number;
Bin = Binarity \{s = single, pb = possible binary, b = probable binary \}; 
Memb = Pleiades Membership Class \{best = very likely member, ok = likely member, nm = not member \}; 
Cdnc = Observing cadence \{LC = long cadence $\simeq$ 30 min, SC = short cadence $\simeq$ 1 min \}; 
$\log \sigma_\text{H}$ = $\log$ total harmonic RMS (ppm); 
$\log \sigma_\text{C}$ = $\log$ continuum RMS (ppm);
$\alpha_1$ = MFC power law index}
\end{deluxetable}
\end{rotatetable}

\end{document}